\documentstyle[11pt,aaspp4]{article}

\lefthead{Boreiko and Betz}
\righthead{63 $\mu$m \ion{O}{1} Line in M42}

\begin{document}
\def\twcii{$^{\rm 12}$\ion{C}{2}}
\def\thcii{$^{\rm 13}$\ion{C}{2}}
\def\cii{\ion{C}{2}\ }
\def\oi{\ion{O}{1}\ }
\def\kms{\hbox{km\, s$^{\rm -1}$}}
\def\cmtwo{\hbox{cm$^{\rm -2}$}}
\def\cmthree{\hbox{cm$^{\rm -3}$}}
\def\vlsr{V$_{\rm LSR}$\ }
\def\tr{T$_{r}^{*}$}
\def\tkin{T$_{\rm kin}$}
\def\threepone{\hbox{$^{3}P _{1}$\, -\,$^{3}P _{2}$}}
\def\threepzero{\hbox{$^{3}P _{0}$\, -\,$^{3}P _{1}$}}
\def\tonec{$\theta ^{1} $C}

\title{Heterodyne Spectroscopy of the 63 ${\bf \mu}$m {\bf \ion{O}{1}} Line in M42}

\author{R. T. Boreiko and A. L. Betz}
\affil{Center for Astrophysics and Space Astronomy, University of Colorado,
    Boulder, CO  80309}

\begin{abstract}
We have used a laser heterodyne spectrometer to resolve the emission
line profile of the 63 \micron\ \threepone\ fine-structure transition
of \oi at two locations in M42.
Comparison of the peak antenna temperature with that of the
158 \micron\ \cii fine-structure line shows that the gas kinetic temperature
in the photodissociation region near \tonec\ is 175 - 220 K, the density is
greater than ${\rm 2 \times 10 ^{\rm 5}}$\ \cmthree , and the hydrogen
column density is about ${\rm 1.5 \times 10 ^{\rm 22}}$\ \cmtwo .
A somewhat lower temperature and column density are found in the IRc2 region,
most likely reflecting the smaller UV flux. The observed width of the
\oi line is 6.8 \kms\ (FWHM) at \tonec , which is slightly broadened over
the intrinsic linewidth by optical depth effects.
No significant other differences between the \oi
and \cii line profiles are seen, which shows that the narrow emission from
both neutral atomic oxygen and ionized carbon
comes from the PDR.  The \oi data do not rule out the possibility of
weak broad-velocity emission from shock-excited gas at IRc2, but the \cii
data show no such effect,
as expected from non-ionizing shock models.
\end{abstract}


\section{Introduction}

Photodissociation regions (PDR) occur at the interface between \ion{H}{2}\
regions and cooler molecular material.  They are most easily studied via
the fine-structure line emission from 
C$^{\rm +}$ and O and the rotational lines of the abundant CO molecule.
Of particular interest are the fine-structure lines of \oi at
63 \micron\ and \cii at 158 \micron , since these are believed to be
the major cooling lines, with \oi emission usually dominant at densities
typical of dense molecular clouds (${\rm n \, > \, 10 ^{\rm 4}}$ \cmthree ) 
and \cii emission more significant at lower densities (\cite{th85a}).
Since the ionization potential of O is very similar to that of H, essentially
all the oxygen in a PDR will be neutral.  Similarly, since the first
ionization potential of C is similar to the dissociation energy of CO
while its second ionization potential is close to that of He, carbon
throughout the PDR will be singly ionized.  Observations of both major
fine-structure cooling lines from the same region provide information
on the temperature, density, and relative abundance of the species
within the PDR.

Both the 63 \micron\ \threepone\ \oi and the
158 \micron\ $^{2}P_{\case{3}{2}}$\, -\,$^{2}P_{\case{1}{2}}$ \cii lines
have been previously observed and
mapped in Orion (e.g. \cite{mgh79,swt79,wer84,cra86,sta93}).
While comparison of integrated intensities and an assumed linewidth for
both transitions allowed these authors to make the first estimates of
physical conditions,
a definitive determination was hampered by the inability
to resolve line profiles.  The 5 \kms\ (FWHM) linewidth of the
\cii line near \tonec\ in M42 was first measured
by Boreiko, Betz, and Zmuidzinas (1988), 
who examined the dynamics of the photoionized gas.
This Letter describes observations of the 63 \micron\
\threepone\ line of \oi in the Orion region with 0.2 \kms\ resolution,
which gives the first fully resolved line profiles for M42 at this wavelength.
Comparison with 158 \micron\ \cii line profiles at similarly high
resolution allows the determination of the optical depth in both these
lines, and hence the temperature and limits to the density within the
PDR.

\section{Instrumentation and Calibration}
The data were obtained using a far infrared heterodyne receiver 
(\cite{bb93}) flown aboard the Kuiper Airborne Observatory (KAO).
The local oscillator (LO) is the 4751.3409 GHz transition
of $^{\rm 13}$CH$_{\rm 3}$OH (\cite{hp78}), which is 6.6 GHz away from 
the rest frequency of the \oi line at 4744.777 GHz (\cite{oifreq}).
The mixer is a GaAs Schottky diode (University of Virginia type 1T15)
operated at room temperature in a corner-reflector mount.  The system
noise temperature measured during the observations was 140,000 K (SSB).
The IF signal is analyzed by a 400-channel acousto-optic spectrometer (AOS) with
a channel resolution of 3.2 MHz (0.2 \kms ) and bandwidth of 80 \kms\ at the
\oi frequency. The 158 \micron\ \cii spectra used for comparison with the
\oi data were taken in 1991 November with a system noise temperature of
9500 K (SSB) and a chopper throw of 12\farcm 5 . 
Other technical details are similar to
those presented by Boreiko et al. (1988), with the addition of the AOS
for signal analysis.

The \oi observations were done using sky shopping at 4 Hz, with an
amplitude of 7\farcm 5 - 8\arcmin , which is sufficient to ensure negligible
emission in the reference beam (see the map of \cite{wer84}).  The
telescope beam size was 17\arcsec\ (FWHM) and the pointing accuracy
is estimated to be $\sim$15\arcsec . Absolute intensity calibration
is derived from spectra of the Moon, for which a physical temperature of
394 K and emissivity of 0.98 are adopted (\cite{lin73}).  Conversion from
DSB to SSB intensity was performed using the calculated transmission of
the pressure window and the atmosphere in the two sidebands,
corrected for the different source and lunar elevation angles.  The net
calibration uncertainty for a source filling the beam is $\leq$15\% .
The velocity scale accuracy, determined from the known line and LO
rest frequencies, is better than $\pm$0.15 \kms\ (1$\sigma$).

\section{Observations}

The \oi line was observed from the KAO flying at an altitude of 12.5 km on
the nights of 1995 Sept 9 and Sept 11.  Data were obtained at two
locations within the M42 complex: \tonec\ and IRc2 .  The Earth's
orbital velocity and the intrinsic source \vlsr\ combined to produce a
net observational Doppler shift near 2 \kms , so that atomic oxygen in
the Earth's thermosphere produced an optically thick absorption line
within the Orion spectra.  The terrestrial line, however, is very narrow,
with a velocity width of 1.3 \kms\ (FWHM) as determined from the lunar
spectra.  The Orion data were corrected for the absorption over the
range of $>$50\% transmission, while the remaining few affected channels
were removed from the spectra.
The interfering telluric line does not compromise the calibration
accuracy or the determination of the emission line profile,
since the telluric absorption
profile was well measured from the lunar spectra.
Also, it affects fewer than 10 of the approximately 70 resolution elements
across the emission line in Orion (see Fig. 1).

Table 1 presents various parameters of the observed spectra.  The continuum
level is obtained from an average over all the spectral channels excluding
the emission line.  The peak \oi antenna temperature \tr\ (as defined by
Kutner and Ulich 1981) is the average value
of the strongest 1.0 \kms\ for \tonec\ and 0.6 \kms\ for IRc2, while that
for \cii is the peak measured single-channel value.
The widths are measured from the raw data since the line shapes for both
\oi and \cii are not Gaussian to within statistical uncertainty.  
Velocity limits for the integrated
intensity are $\Delta$v(FWHM) on either side of line center.
\placetable{tbl1}

The measured \oi continuum given in Table 1 is in good agreement with
that obtained from a map by Werner et al. (1976) at \tonec , but is a factor of
2.5 higher at IRc2.  The most likely explanation for the apparent 
discrepancy is the different beamsizes of the observations, 17\arcsec\  here
vs 1\arcmin\ for the data of Werner et al. (1976).  A continuum source size
of $\sim$40\arcsec\ FWHM,
similar to that seen in 119 \micron\ continuum data (\cite{bb89}), would
reconcile the measured values.  The integrated intensity values of Table 1
are in good agreement with those reported previously 
(\cite{swt79,wer84,cra86}).

\section{Analysis and Interpretation}

\subsection{Gas Kinetic Temperature}

\begin{figure}
\centering
\leavevmode
\epsfysize=7.2in
\epsfbox{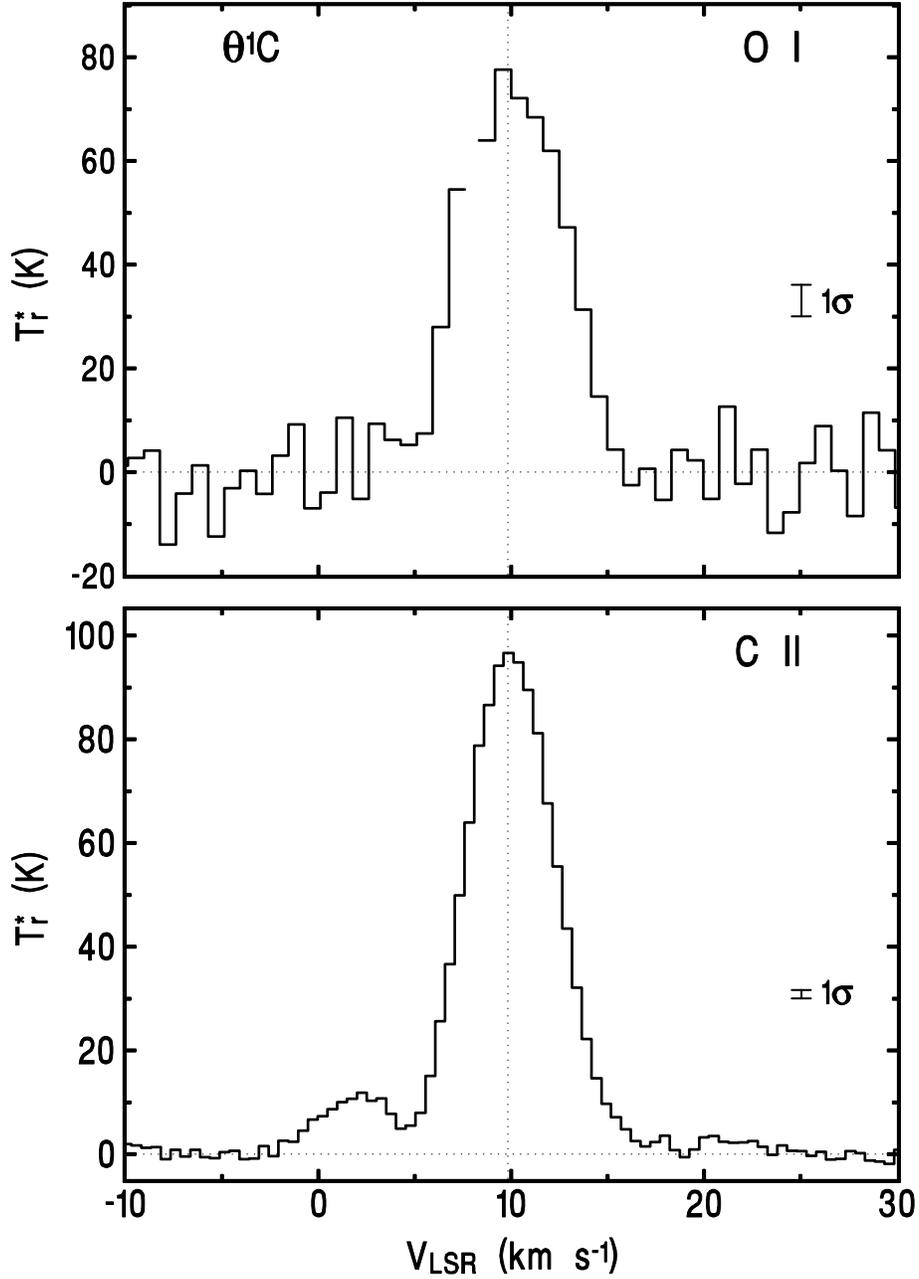}
\caption[fig1.eps]{Observed spectra of the 63 \micron\ O\, {\sc i}\ and 
158 \micron\ C\, {\sc ii}\ fine-structure lines toward \tonec . 
Integration times are 72 minutes for O\, {\sc i}\ and 12 minutes for 
C\, {\sc ii}.  The continuum has been removed from both spectra.
The light vertical line is at \vlsr $=$9.8 \kms .
The gap in the O\, {\sc i}\ spectrum near \vlsr$=$8 \kms\ shows the
region where telluric atomic oxygen absorption exceeds 50\%.
\label{fig1}}
\end{figure}

Figure 1 presents spectra of the \threepone\ \oi and the \cii 
$^{2}P_{\case{3}{2}}$\, -\,$^{2}P_{\case{1}{2}}$
fine-structure lines taken with the same instrument toward \tonec .
A lower limit to the gas kinetic temperature of 165 K is given by the peak
brightness temperature of the \oi line.  However, because of the high
critical density of the \threepone\ transition (n$_{\rm c} \sim$5$\times$10
$^{\rm 5}$
\cmthree ), the gas temperature may be higher unless the PDR has 
sufficient
density or the line has significant optical depth.  The \cii transition has
a much lower critical density, n$_{\rm c} \sim$3$\times$10$^{\rm 3}$ \cmthree ,
and thus is generally in LTE.  Comparison of just the two peak brightness
temperatures
shows that the \cii line must have an optical depth $\tau <$1.6.
\placefigure{fig1}

\begin{figure}
\centering
\leavevmode
\epsfysize=7.2in
\epsfbox{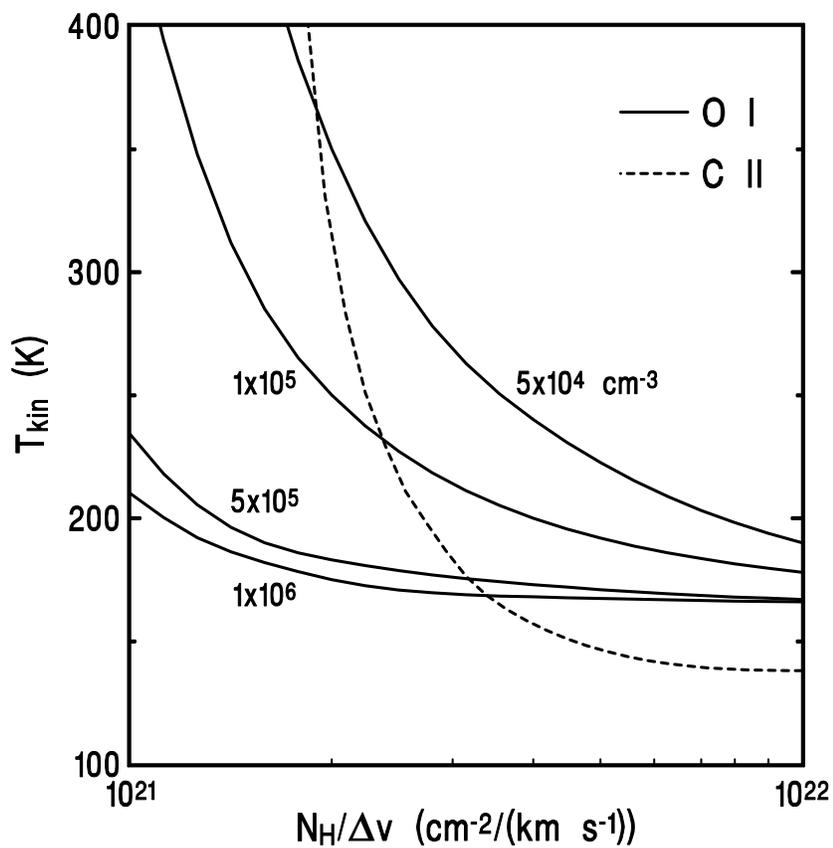}
\caption[fig2.eps]{Combinations of gas kinetic temperature and hydrogen 
column density per unit velocity interval which produce the observed 
O\, {\sc i}\ 63 \micron\ (solid lines) and C\, {\sc ii}\ 158 \micron\ 
(dashed line) peak antenna temperatures at \tonec .  
The O\, {\sc i}\ curves are calculated for various gas densities.
See text for details.  \label{fig2}}
\end{figure}

Figure 2 presents curves showing the combinations of gas kinetic temperature,
density, and column density of hydrogen which produce the observed peak
antenna temperatures for the 63 \micron\ \oi and 158 \micron\ \cii lines
at \tonec .  The horizontal axis gives column density of hydrogen per
unit velocity interval in a LVG model.  Relative abundances of O and C
are assumed to be 5$\times$10$^{\rm -4}$ and 3$\times$10$^{\rm -4}$,
respectively.  The rate coefficients of Launay and Roueff (1977) and A
coefficients of Fischer and Saha (1983) for \oi were used in a statistical
equilibrium escape probability formalism.  Only an LTE solution is shown
for the \cii curve.
As can be seen from the figure, consistency with both
observations requires that the PDR gas temperature be less than 250 K for
moderate densities, n$_{\rm H} > {\rm 1 \times 10 ^{\rm 5}}$ \cmthree .
An additional constraint to the allowed range of temperature and density
is provided by the 145 \micron\ \threepzero\ \oi line.
Although fully resolved spectra for this higher excitation line are not
available, its integrated intensity has been measured to be 
(4.0$\pm$1.3)$\times$10$^{\rm -3}$
${\rm erg \, cm ^{\rm -2} \, s ^{\rm -1} \, sr ^{\rm -1}}$ at \tonec\
(as quoted by
\cite{sta93}; a 30\% uncertainty has been assumed).  This relatively high 
value by itself excludes all potential solutions from Fig. 2 for which
n$<$1$\times$10$^{\rm 5}$ \cmthree , since the critical density for this
transition is 6$\times$10$^{\rm 4}$ \cmthree .
Although no combination of parameters can reproduce all the observed
intensities simultaneously, the best agreement is found at the higher
densities, n$\geq$5$\times$10$^{\rm 5} \ {\rm cm}^{\rm -3}$, where the
63 \micron\ \oi line has moderate optical depth.  
The lowest density for which a consistent solution 
is possible within calibration uncertainties and 1$\sigma$ statistical
uncertainty for all the data is 
${\rm n _{\rm H} = 2 \times }$10$^{\rm 5}$ \cmthree ,
with \tkin$=$220 K and 
${\rm N _{\rm H} / \Delta v = 2.8 \times 10 ^{\rm 21} \ cm ^{\rm -2} \,
{(km \, s^{\rm -1})} ^{\rm -1}}$.  Densities of n$_{\rm H} > {\rm 10}^{\rm 6}$
\cmthree\ such that the \oi lines are thermalized are also consistent with
all the data to within statistical and calibration uncertainties.  Thus,
with no further constraints, the optical depth in the \oi 63 \micron\
line is between 2.6 and 5.3, while that for the 158 \micron\ \cii line is
1.0 - 1.3.  If the abundance of oxygen relative to carbon is higher than
the 5:3 ratio assumed for the calculations, then the 63 \micron\ line optical
depth would be increased and a brighter 145 \micron\ \oi line would be
predicted.  However, this would not change the basic conclusion that 
the PDR near \tonec\ is characterized by
\tkin$\, \leq \,$220\ K, ${\rm n _{\rm H} \, \geq \,
2 \times 10 ^{\rm 5}}$ \cmthree , and ${\rm N _{\rm H} \, \sim \,
1.5 \times 10 ^{\rm 22}}$ \cmtwo . Additional constraints on these
parameters are presented in section~\ref{optd}.
\placefigure{fig2}

Equivalent calculations at the IRc2 position have the same difficulty
reproducing the reported high 145 \micron\ \oi integrated intensity
simultaneously
with the low peak 63 \micron\ \oi antenna temperature.  It should be
noted that calibration of the 145 \micron\ emission is 
difficult because it lies on the edge of a strong atmospheric feature.
Again, the best
solutions are found at the lower temperatures, 140\ K $\leq$ \tkin $\leq$
180\ K, with densities ${\rm n _{\rm H} \, \geq \, 1 \times 10 ^{\rm 5}}$ 
\cmthree , and ${\rm n _{\rm H} / \Delta v  \, = \, (1.9 \, - \, 2.7)
\times 10 ^{\rm 21} \ cm ^{\rm -2} \, {(km \, s^{\rm -1})} ^{\rm -1}}$.  
The higher densities and column densities are associated with lower
kinetic temperatures.  The optical depth in the 63 \micron\ \oi line
lies between 3.6 and 5.1, while that in the 158 \micron\ \cii line is 
0.8 - 1.4 .

\subsection{Line Width and Optical Depth} \label{optd}

The \cii line at both observed locations is narrower than the \oi line
by 25-40\%.  If the differences were due to the different beamsizes (44\arcsec\ 
for \cii vs 17\arcsec\ for \oi) or critical densities for the two lines,
one would generally expect the opposite effect.  
Therefore it is reasonable to assume
that the major cause of the greater \oi line width is a larger optical depth.
For $\tau _{\rm C \, II}$ between 1.0 and 1.3 
(the range appropriate to the
solutions at \tonec\ discussed above), the relative linewidths of the
two transitions suggest that $\tau _{\rm O \, I,\ 63\, \mu m }$ is
\hbox{3 - 4}, if a Gaussian atomic velocity distribution 
is assumed.  At IRc2, the calculated optical depth of the 63 \micron\ \oi
line is somewhat larger, 4 - 5.5.  These \oi optical depths are in good
agreement with those calculated from the peak antenna temperatures of the
\oi and \cii lines throughout the range of temperatures derived above.

The intrinsic linewidth in the PDR is small, $\sim$4.5 \kms\ (FWHM)
near \tonec\ and $\sim$3.6 \kms\ at IRc2, after correction for the
optical depth in the fine-structure lines.  This value agrees well with
that seen in carbon recombination lines in the same region (4.4 \kms ;
\cite{jp78}).  We see no evidence for a 
broad velocity component at \tonec , although as can be seen from Fig. 1, there
is an additional velocity component near 3 \kms , at least in the 158
\micron\ \cii data.  At IRc2, both the \cii and \oi lines are asymmetric,
which indicates a contribution to the emission from material associated
with the molecular ridge at a velocity near 10 \kms\ in addition to that
from photodissociated gas at \vlsr $\sim$ 8 \kms . An upper limit
(1$\sigma$) to the antenna temperature (\tr ) of a broad emission component
($\Delta \rm v _{\rm FWHM} \, = \,$25 - 40 \kms ) is 6 K for \ion{O}{1}.
Any broad emission, presumably from shock-excited gas,
could contain an integrated intensity equivalent to
that in the narrow component (listed in Table 1), as suggested by the
data of Crawford et al. (1986) and Werner et al. (1984).  However,
the measured line width (FWHM) remains unaffected by the broad component.  

\subsection{Comparison with PDR Models and Previous Measurements}

The PDR model of Tielens and Hollenbach (1985a, 1985b) solves equations
of chemical equilibrium and energy balance to calculate temperature and
number density profiles through a photodissociation region illuminated by
UV radiation.  Their best-fit model for the \tonec\ region has densities
${\rm n _{\rm H} \, = \, 1 \times 10 ^{\rm 5}}$ \cmthree\ and a peak
temperature near 550\ K.  The parameters, however, are determined from
integrated intensity measurements of \oi and \cii emission and assumed
identical line widths of 4.5 \kms\ (FWHM),
which are smaller than the true (and different) values for \oi
and \ion{C}{2}, as
can be seen from Table 1 and Fig. 1.  Thus the gas kinetic
temperature of 265 K calculated by Tielens and Hollenbach (1985b)
for the moderately optically thick \oi line is too high.

There are several advantages to determining temperature and density
from the peak emission of spectrally well-resolved lines rather than from
the integrated intensity.
There is no confusion from different velocity
components, which can affect comparisons of integrated intensity.  For
example, Fig. 1 shows a 3 \kms\ component prominently in the \cii data,
but only weakly (if at all) in the \oi 63 \micron\ spectrum.
This feature contributes 10\% to the total integrated intensity in
the \cii line but is clearly irrelevant to the determination of
physical parameters of the main 10 \kms\ component of the PDR. 
Inclusion of this component in an
assumed 5 \kms\ linewidth leads to a brightness temperature
T$_{\rm B} \sim$165 K for \cii , as calculated by Stacey et al.\ (1993)
rather than 138 K, as seen in Table 1.  The difference is not very
significant in this case, but other locations within the Orion complex
show line profiles with large asymmetries and additional velocity
components.

The effect of reference beam emission is less severe for peak temperature
measurements than for integrated intensity if velocity differences exist
between the desired and potentially contaminating emission.
Even with no velocity gradient, the 12\farcm 5 chop throw for the \cii
observations limits the reduction of our peak \cii \tr\ to less than
10\%, calculated from the strip scan of
Stacey et al. (1993).
The good agreement between our measured integrated intensity and that shown
in this strip scan (which is corrected for self-chopping) also suggests
that there was little emission in our reference beams.
No reduction of the peak \oi \tr\ is likely from reference beam emission,
as shown by the \oi map of Werner et al.\ (1984).
Thus our conclusions remain substantially unchanged.

Finally, a single linewidth does not need to be assumed for all spectral
features if they are resolved, so that line broadening from optical depth
effects does not distort estimates of brightness temperature.  Rather, the line
width can be used as supporting evidence for the derived optical depth,
as discussed in section~\ref{optd}.  

Stacey et al. (1993) constructed a model for the PDR near \tonec\ based
on 63 \micron\ \oi and 158 \micron\ \cii integrated intensity measurements.
This model is characterized by a kinetic temperature \tkin $\sim $300\ K,
density n$_{\rm H} \sim$4$\times$10$^{\rm 5}$ \cmthree , and column density
N$_{\rm H} \sim$1.5$\times$10$^{\rm 22}$ \cmtwo .
Since the \oi and \cii integrated intensities of the present data agree
well with those used by Stacey et al.\ (1993), it is to be expected that
those derived parameters which are independent of line width are also very
similar in the two models.
However, the 300 K derived kinetic temperature of Stacey et al. (1993) is
significantly higher than ours (\tkin$=$175\,-\,220\ K) because they
assumed the linewidth of the 63 \micron\ \oi line to be 5 \kms ,
similar to that of the \ion{C}{2}, while the true value is near
6.8 \kms\ because of optical depth broadening, as seen in the present data.
Spectrally resolved observations of the
145 \micron\ \threepzero\ \oi line should further constrain estimates of the 
temperature and density of the Orion PDR.

The peak optical depth of the 158 \micron\ \cii line of $\sim$0.9 calculated
by Stacey et al. (1993) is lower than that obtained from the present data
because of the difference in model kinetic temperatures, while the 63
\micron\ \oi line has comparable model optical depths.  
From comparison
of the integrated intensity of the F$=$1-0 hyperfine component of the
158 \micron\ \thcii\ fine-structure line with that of the 
\twcii\ line, Stacey et al. (1991) obtained a relationship between 
the line-averaged \twcii\ optical depth and the $^{\rm 12}$C/$^{\rm 13}$C
isotopic ratio R :
\[ \frac{\tau }{1 - e ^{- \tau }} = 1.45 (^{+0.34}_{-0.24}) \cdot
\frac{R}{43} 
\] at \tonec .
For their assumed isotopic ratio of 43, the corresponding peak 
optical depth is 1.2${\rm (^{\rm +0.7}_{\rm -0.6} ) }$, in agreement with the 
range of 1.0\,-\,1.3 derived in this paper.

\acknowledgments

We thank the staff of the Kuiper Airborne Observatory for their consistent
support, especially during the stressful last month of KAO operations.
This work is supported by NASA grant NAG 2-753.

\clearpage
\begin{deluxetable}{lcclrcc}
\footnotesize
\tablecolumns{7}
\tablewidth{0pc}
\tablecaption{Observed O\, {\sc i}\ and C\, {\sc ii}\ Line Parameters\tablenotemark{a}\label{tbl1}}
\tablehead{
\colhead{Position} & \colhead{V$_{\rm LSR}$} & \colhead{Linewidth} & \colhead{Peak T$_{r}^{*}$\ \ \ \ }& 
\colhead{\ \ \ \ T$_{\rm B}$} & \colhead{Integrated Intensity} & \colhead{Continuum T$_{r}^{*}$} \nl 
\colhead{}         & \colhead{(km\, s$^{\rm -1}$)}  & \colhead{(km\, s$^{\rm -1}$, FWHM)}  & \colhead{(K)\ \ \ \ }  &
\colhead{\ \ \ \ (K)} & \colhead{(erg\, cm$^{\rm -2}$\, s$^{\rm -1}$\, sr$^{\rm -1}$)}   & \colhead{(K)}	}
\startdata
\cutinhead{63 \micron\ O\, {\sc i}\ line\tablenotemark{b}}
$\theta _{1}$C & 9.8(0.2) & 6.8(0.4) & 76.8(5.1) & 165(6) & 5.42(0.15)$\times$10$^{\rm -2}$ & 0.4(0.3) \nl
IRc2 & 8.5(0.2) & 5.8(0.4) & 54.7(6.4) & 139(8) & 3.15(0.14)$\times$10$^{\rm -2}$ & 4.2(0.3) \nl
\cutinhead{158 \micron\ C\, {\sc ii}\ line\tablenotemark{c}}
$\theta _{1}$C & 9.9(0.1) & 5.4(0.3) & 97.4(1.6) & 138(2) & 4.32(0.03)$\times$10$^{\rm -3}$ & 1.0(0.1) \nl
IRc2 & 8.4(0.2) & 4.2(0.3) & 74.6(0.6) & 114(1) & 3.00(0.02)$\times$10$^{\rm -3}$ & 5.0(0.1) \nl
\enddata
\tablenotetext{a}{Numbers in parentheses represent 1$\sigma$ uncertainties.}
\tablenotetext{b}{narrow component only}
\tablenotetext{c}{Integrated intensity includes several velocity components.}
\end{deluxetable}

\clearpage

% Now comes the reference list.  

\clearpage

\end{document}